\documentclass[slac_one]{revtex4}
\usepackage{graphicx}
\usepackage{fancyhdr}
\usepackage{xspace}
\def\TeV{\ifmmode {\mathrm{\ Te\kern -0.1em V}}\else
                   \textrm{Te\kern -0.1em V}\fi \xspace}%
\def\GeV{\ifmmode {\mathrm{\ Ge\kern -0.1em V}}\else
                   \textrm{Ge\kern -0.1em V}\fi \xspace}%
\def\MeV{\ifmmode {\mathrm{\ Me\kern -0.1em V}}\else
                   \textrm{Me\kern -0.1em V}\fi \xspace}%
\def\ifb{\mbox{fb$^{-1}$} \xspace}

\begin{document}

\title{{\small{Hadron Collider Physics Symposium (HCP2008),
Galena, Illinois, USA}}\\ 
\vspace{12pt}
Higgs Boson Properties and BSM Higgs Boson Searches at LHC} 

\author{W.~F.~Mader (on behalf of the ATLAS and CMS collaborations)}
\affiliation{ Technical University Dresden, 01062 Dresden, Germany}
\begin{abstract}
At the end of 2008, the Large Hadron Collider (LHC) will come into
operation and the two experiments ATLAS and CMS will start taking data
from proton-proton collisions at a center-of-mass energy of
$\sqrt{s}=14\,\TeV$. In preparation for the data taking period, the
discovery potential for Higgs bosons beyond the Standard Model
has been updated by both experiments and is reviewed here. In
addition, the prospects for measuring the properties of a Higgs boson
like its mass and width, its CP eigenvalues and its couplings to
fermions and gauge bosons are discussed. 
\end{abstract}

\maketitle

\thispagestyle{fancy}

\section{MSSM HIGGS BOSON SEARCHES}
In the Minimal Supersymmetric Standard Model (MSSM), the minimal
extension of the Standard Model (SM), two Higgs
doublets are required, resulting in five observable Higgs
bosons. Three of them are electrically neutral ($h$, $H$,
and $A$) while two of them are charged ($H^\pm$). At tree level their
properties like masses, widths and branching fractions can be
predicted in terms of only two parameters, typically chosen to be the
mass of the CP-odd Higgs boson, $m_A$, and the ratio of
the vacuum expectation values of the two Higgs doublets, $\tan\beta$.

In the MSSM the couplings of the Higgs bosons to fermions and bosons are
different from those in the Standard Model resulting in different
production cross-sections and decay rates. The coupling of the Higgs
bosons to third generation fermions is strongly enhanced for large
regions of the parameter space which determines the search strategies
for such Higgs bosons.

In the following, searches for the neutral and charged Higgs bosons in
the MSSM in the two experiments ATLAS and CMS at the LHC are
described. Unless indicated otherwise, all results will be given in
the $m_h^\mathrm{max}$ scenario \cite{Carena:2002qg}. A detailed
description of the ATLAS and CMS detectors can be found elsewhere
\cite{bib:ATLAS, bib:CMS} 


\subsection{Higgs Boson Searches in $h/H/A\to\mu\mu$}
In the SM the discovery potential for the Higgs
boson in the dimuon final state is limited due to its small branching
fraction and the high backgrounds expected from several SM
processes. However, in the MSSM the decay of the three neutral Higgs
bosons $h$, $H$, and $A$ into a dimuon final state can be strongly
enhanced depending on the value of $\tan\beta$. This channel can
therefore serve as a discovery channel for high values of $\tan\beta$,
or as a tool to exclude large parts of the $m_A$ vs. $\tan\beta$
plane. In addition, in the intense coupling region around
$m_A=130\,\GeV$ where all the neutral Higgs bosons have a comparable
mass, the excellent invariant mass resolution of the dimuon final
state offers the potential to observe these states individually at the
same time. 

The event selection is optimized separately for the cases where zero
r at least one $b$ jet are identified in the final state. In the
first case Drell-Yan $Z$ boson production is the dominant background,
while in the second case the dominant contribution is approximately
equally shared between the Z and $t\bar t$ processes. At the
preselection step two isolated muons of opposite charge are required
inside the acceptance region with a $p_T>20\,\GeV$. Due to the high
momenta of the muons from the Higgs boson decays this final state can
be efficiently triggered by a high-$p_T$ single-muon trigger. Further
selection criteria include an upper cut on the amount of missing
energy found in the event, requirements on the jet identified as
coming from a $b$ quark (or a $b$-jet veto in the case of the zero
$b$-jet analysis), on the accoplanarity of the dimuon system and the
sum of $p_T$ of all jets in the event (in the case of the one $b$-jet
analysis).  

\begin{figure}[!t]
  \includegraphics[width=.8\textwidth]{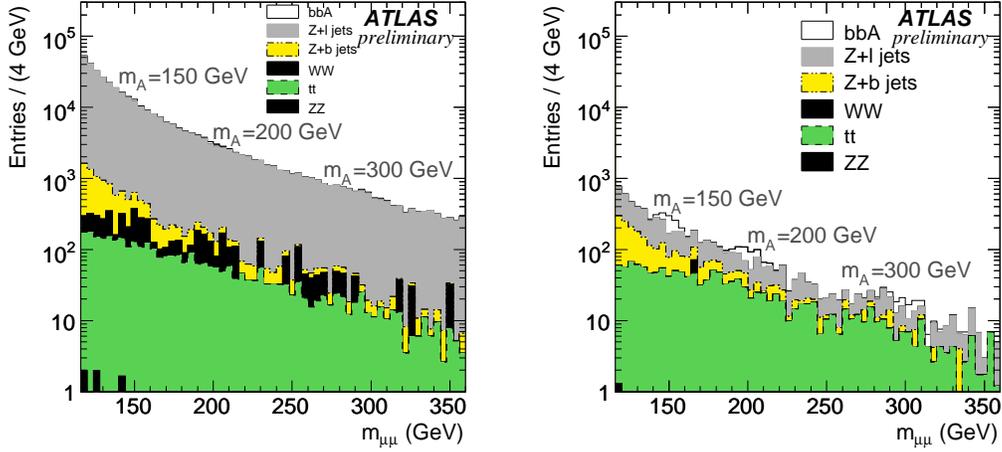}
  \caption{Invariant dimuon mass distribution of the main backgrounds
    and the $A$ boson signal at masses $m_A=150$, $200$, and
    $300\,\GeV$ and $\tan\beta=30$, obtained for an integrated
    luminosity of $30\,\ifb$ in ATLAS. $B$-tagging has been applied for the
    event selection. The production rates of $H$ and $A$ bosons have
    been added. Zero or at least one $b$-jet have been required in the
    left-hand side and the right-hand side plot, respectively.}
  \label{fig:dimuon}
\end{figure}

The invariant dimuon mass after all selection criteria is displayed
for the cases of zero and at least one $b$-jet on the left-hand side
and the right-hand side of Figure \ref{fig:dimuon}, respectively. The 
background from $Z$ and $t\bar t$ events is estimated from data.
The $Z$ background is estimated from data by analyzing the $ee$
final state which is signal-free. In the case of $t\bar t$ events, the
$e\mu$ final state or a $t\bar t$ enriched sample obtained by
requiring large missing $E_T$ in the event can be used.

\begin{figure}[!b]
  \includegraphics[width=.38\textwidth]{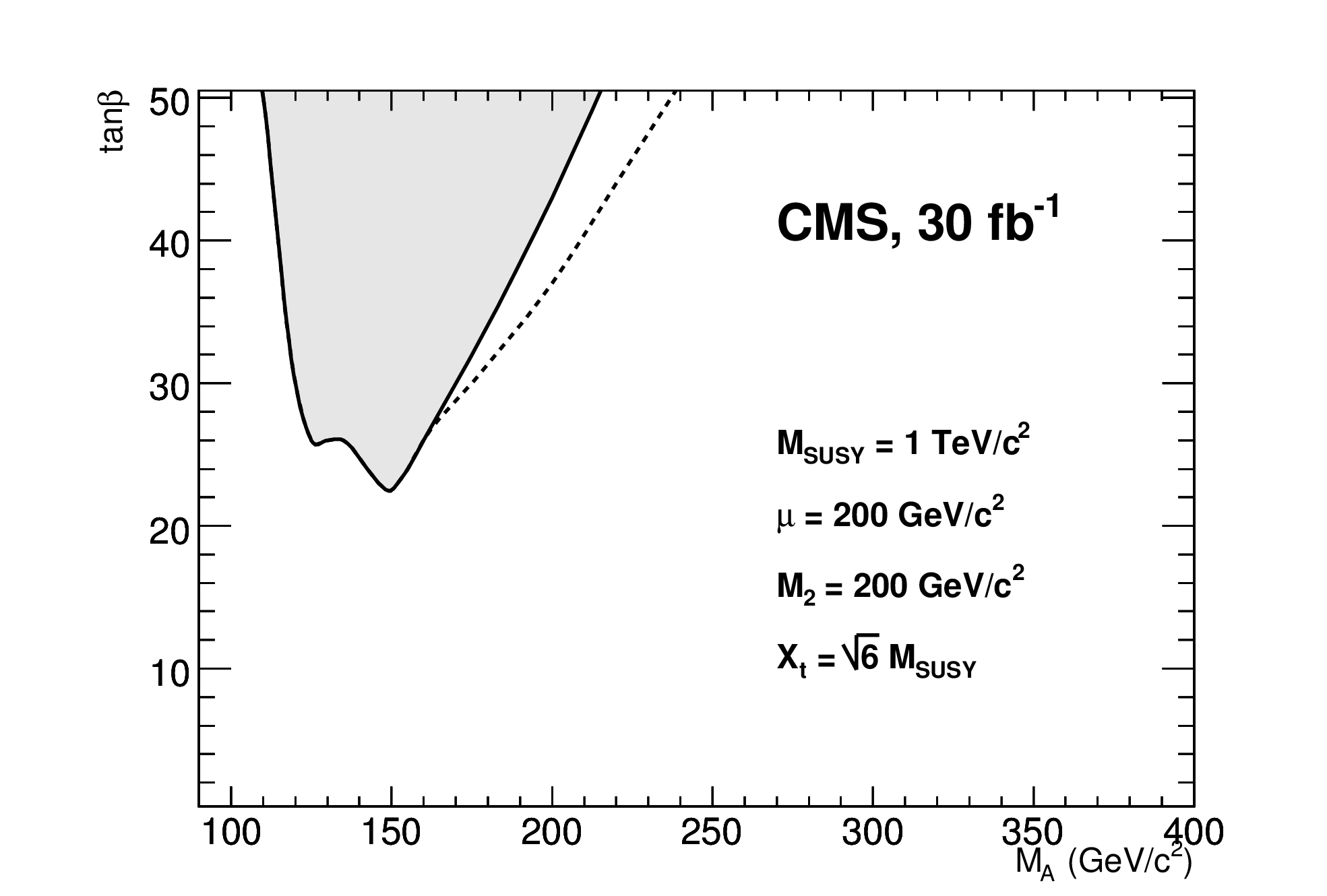}
  \includegraphics[width=.41\textwidth]{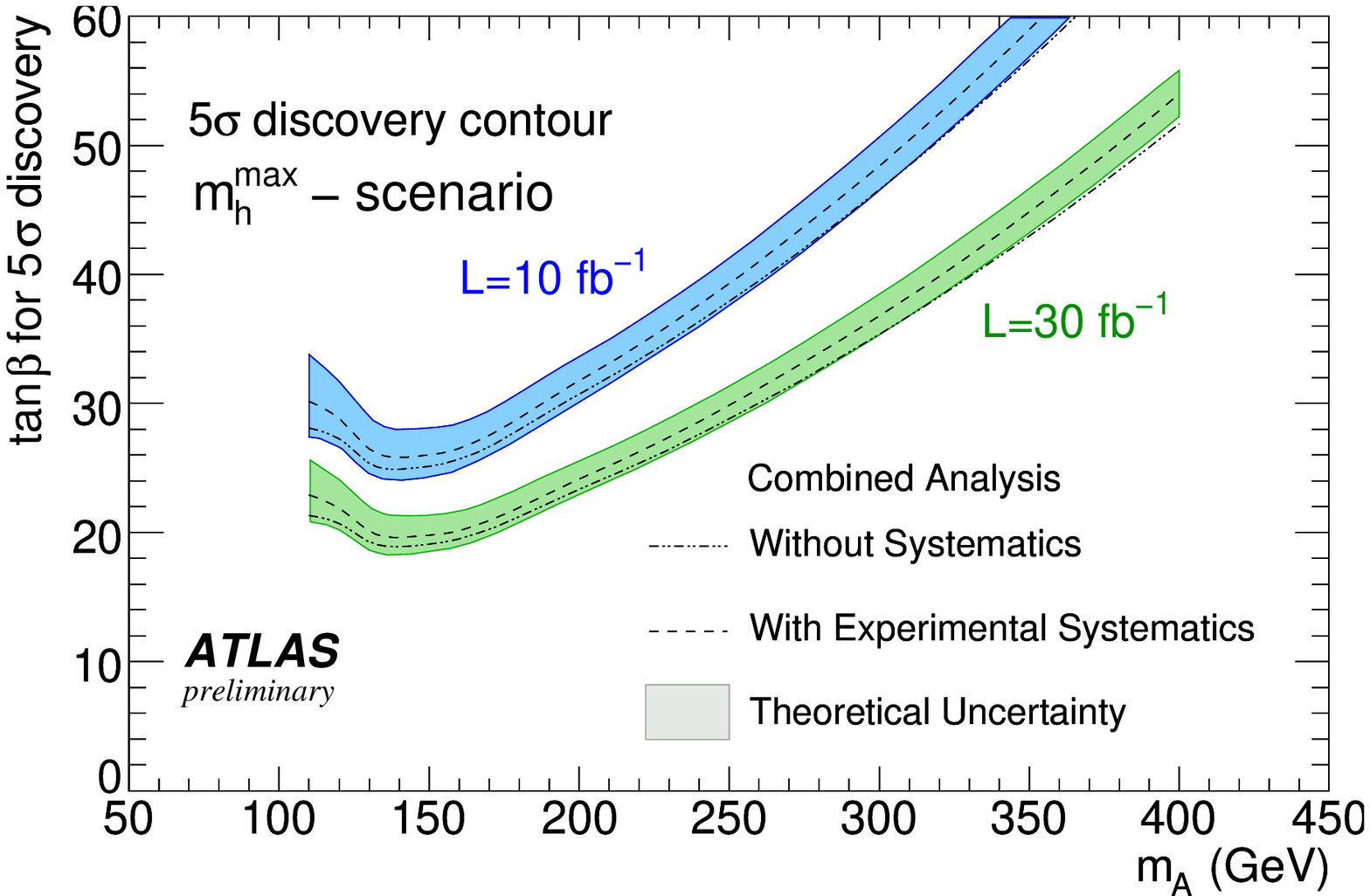}
  \caption{Discovery potential in the $h/H/A\to\mu\mu$ channel as a
    function of $m_A$ and $\tan\beta$. Left: Discovery potential from
    CMS for an integrated luminosity of $30\ifb$. In the shaded area
    $>5\,\sigma$ significance can be obtained; the dashed line
    corresponds to the $5\,\sigma$ contour without systematic
    uncertainties included. Right: The $5\,\sigma$ discovery contours
    from ATLAS corresponding to integrated luminosities of $10$ and
    $30\,\ifb$ with (dashed line) and without (dotted line) systematic
    uncertainties taken into account. The theoretical uncertainties
    are illustrated by the shaded bands.}
  \label{fig:mumuDiscPot}
\end{figure}

The discovery potential as obtained by CMS and ATLAS is displayed
on the left-hand side and the right-hand side of Figure
\ref{fig:mumuDiscPot}, respectively \cite{Ball:2007zza,CSC:Book}. The
theoretical uncertainties include the uncertainty on the
parton density functions and on the renormalization and factorization
scales. The dominant experimental uncertainties (of about $10\%$ in
total) include the reconstruction efficiency, momentum resolution and
momentum scale for the muons, the jet energy scale and resolution as
well as the uncertainty on the $b$ tagging efficiency and the
light-jet rejection rate.

\subsection{Higgs Boson Searches in $h/H/A\to\tau\tau$}
Compared to the dimuon final state, $h/H/A\to\tau\tau$ decays have a
substantially larger branching fraction which scales as
$(m_\tau/m_\mu)^2$ for a given $\tan\beta$. However, since the $\tau$
leptons can decay both, leptonically and hadronically, the signatures
observed in the detector are very different and therefore have to be
treated individually. 


\subsubsection{Higgs Boson Searches in $h/H/A\to\tau\tau\to\ell\ell 4\nu$}
Even though the branching fraction of the two $\tau$ leptons into a fully
leptonic final state is only $12\%$, it contributes
significantly to the discovery potential in particular for Higgs boson
masses in the range $110<m_A<300\,\GeV$. These events can be triggered
on with high efficiency using an electron and/or muon trigger.

The event topology consists of two leptons, missing $E_T$, and
jets. The main background is coming from Drell-Yan $\tau\tau$
production, $t\bar t$ processes, and $W$+jets topologies. The CMS
analysis only studies the $e\mu$ final state \cite{CMS:emu} while
ATLAS exploits all leptonic final states \cite{CSC:Book}. The invariant
di-$\tau$ mass is reconstructed using the collinear approximation
technique \cite{Ellis:1987xu}. 

In Figure \ref{fig:llmass} the invariant di-$\tau$ mass distribution
as obtained in ATLAS is shown for Higgs boson masses of $m_A=110\,\GeV$, and
$m_A=200\,\GeV$, $\tan\beta=20$ and an integrated luminosity of
$30\,\ifb$. For low masses, the irreducible $Z\to\tau\tau$ background
dominates over that from $t\bar t$ processes, while for high masses
the situation is reversed. The dominant experimental systematic
uncertainties come from the jet-energy scale and resolution
uncertainty and from the uncertainty on the $b$-tagging
efficiency. The theoretical uncertainties come from the uncertainties
on the parton density functions, on the factorization and
renormalization scale for signal and $t\bar t$ background. 

The discovery potential for this channel is shown in Figure
\ref{fig:llDiscPot} for ATLAS (left) and CMS (right). For low Higgs
boson masses of the order of $130\,\GeV$, a discovery with at least
$5\,\sigma$ significance will be possible for $\tan\beta$ values of
$15$ or larger, while for higher Higgs boson masses like
$m_A=200\,\GeV$ $\tan\beta$ of the order of $30$ would be necessary.  

\begin{figure}[!t]
  \includegraphics[width=.4\textwidth]{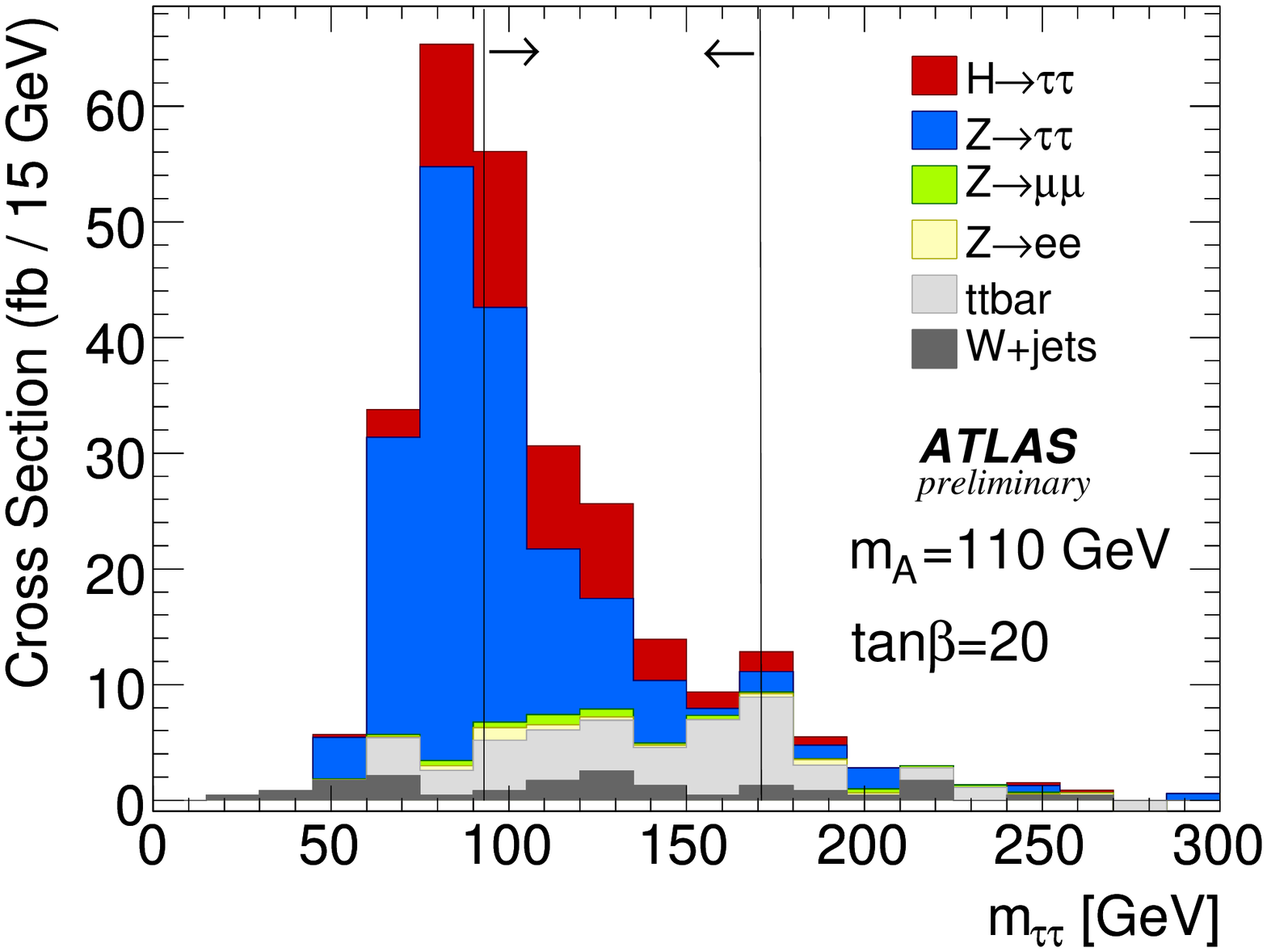}
  \includegraphics[width=.4\textwidth]{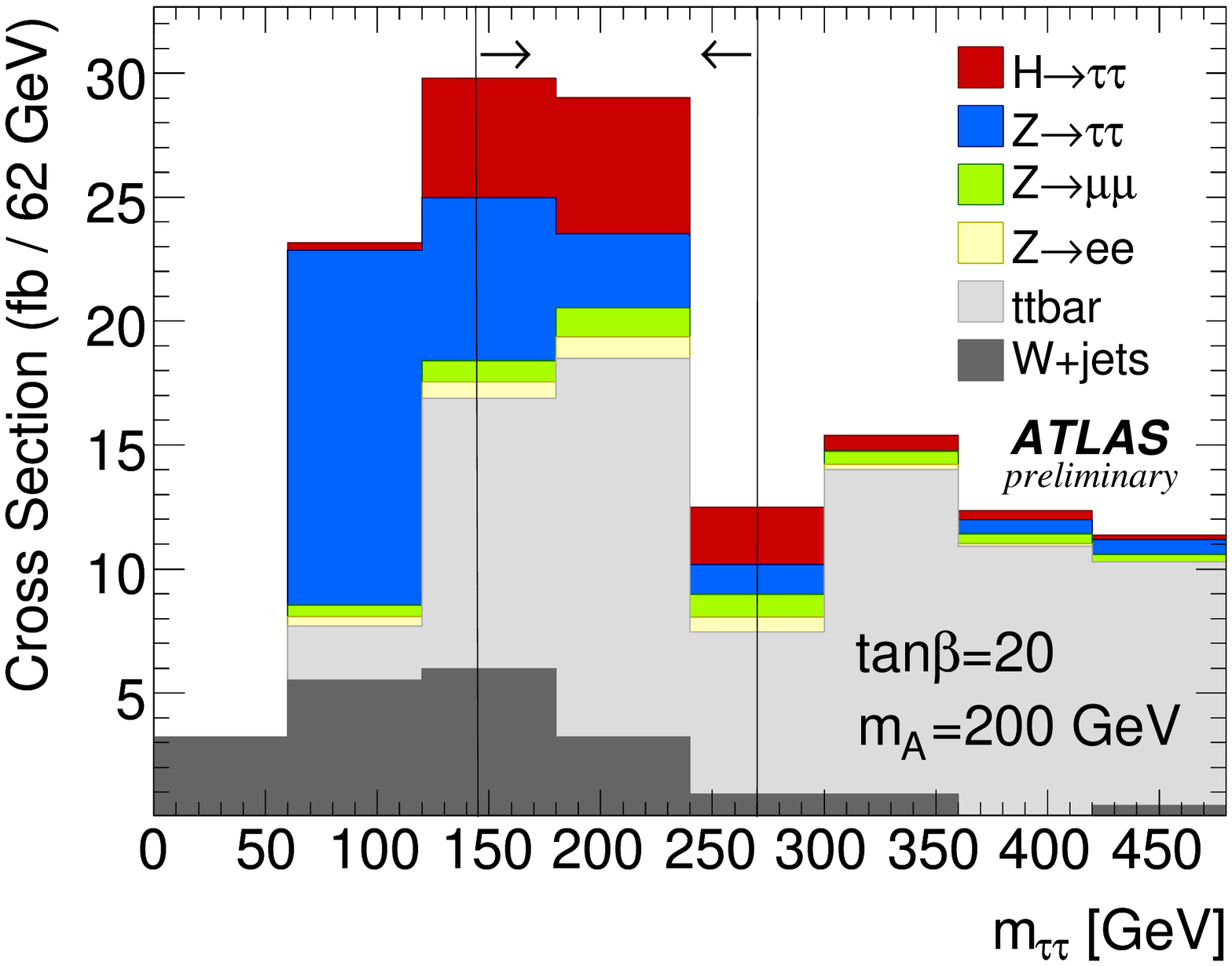}
  \caption{Invariant $m_{\tau\tau}$ distribution for signal and
    background events. The distributions are shown after all selection cuts
    for Higgs bosons masses and $\tan\beta$ values as indicated in the
    plots. The vertical lines indicate the mass window used
    for calculating the signal significance.}
  \label{fig:llmass}
\end{figure}

\begin{figure}[!t]
  \includegraphics[width=.43\textwidth]{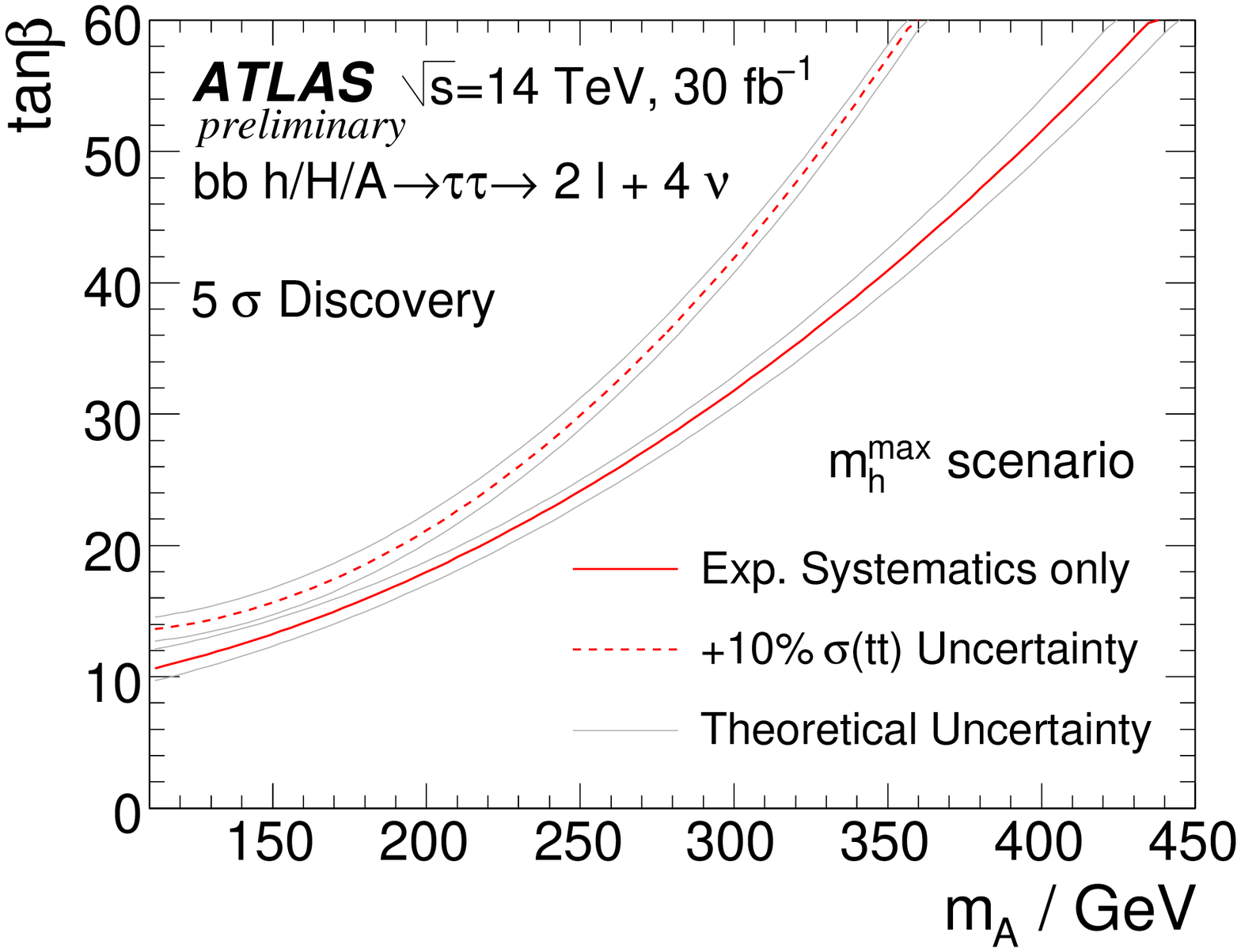}
  \includegraphics[width=.37\textwidth]{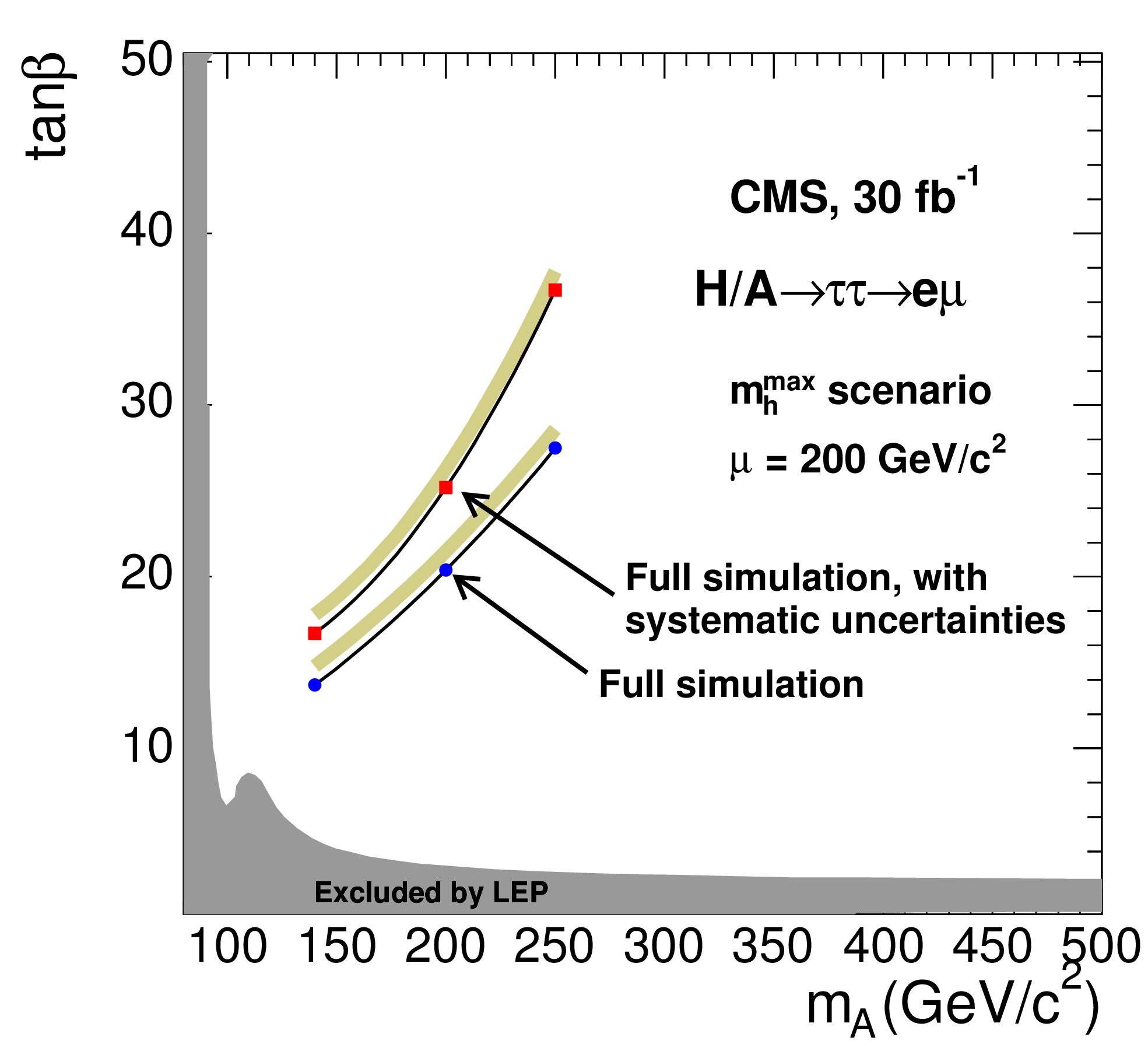}
  \caption{The $5\,\sigma$ discovery contours in the $m_A$
  vs. $\tan\beta$ plane for the fully leptonic final state and for an
  integrated luminosity of $30\,\ifb$ with systematic uncertainties
  taken into account for ATLAS (left) and CMS (right).}
  \label{fig:llDiscPot}
\end{figure}


\subsubsection{Higgs Boson Searches in $h/H/A\to\tau\tau\to\ell h 3\nu$}

\begin{figure}[!b]
  \includegraphics[width=.47\textwidth]{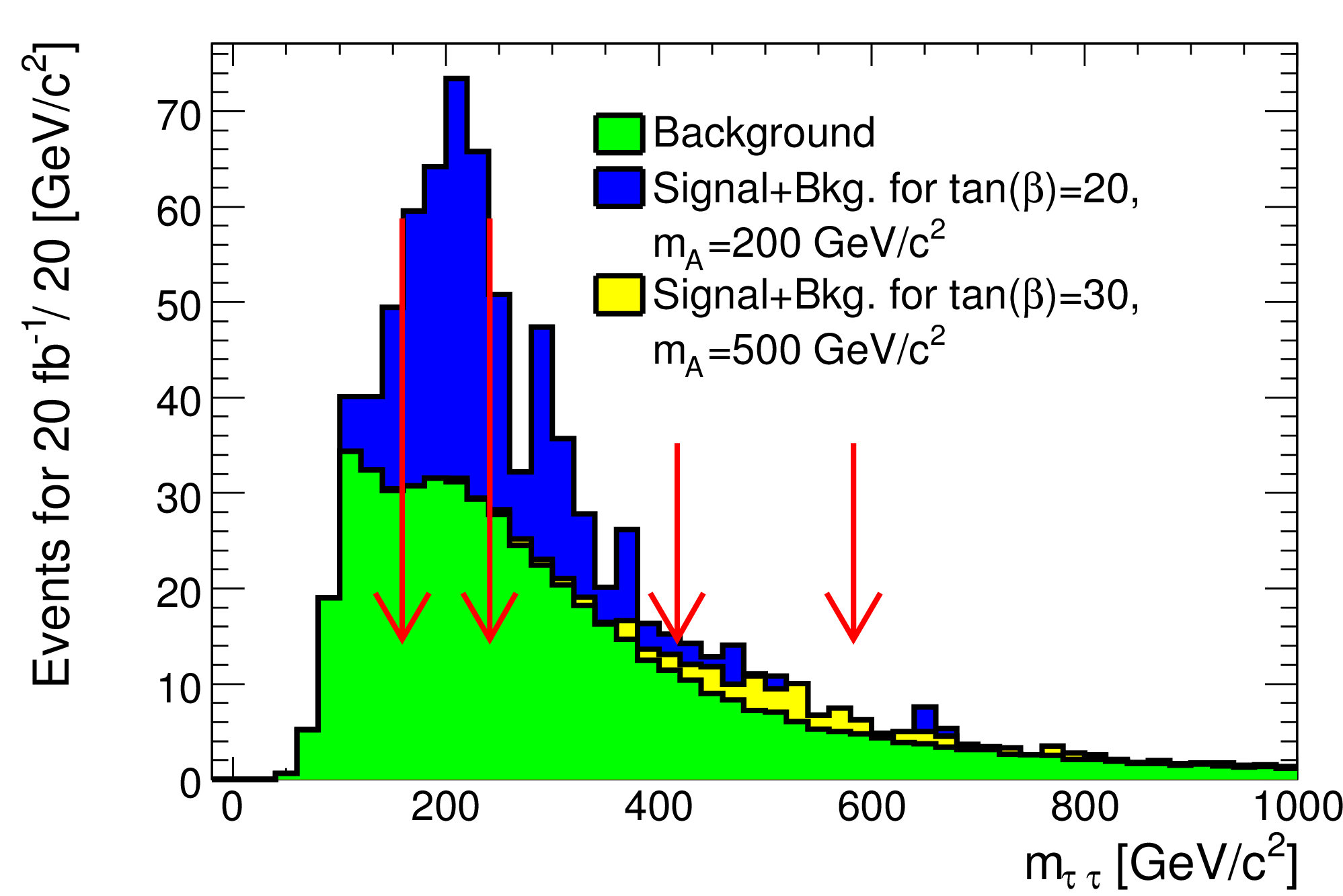}
  \includegraphics[width=.33\textwidth]{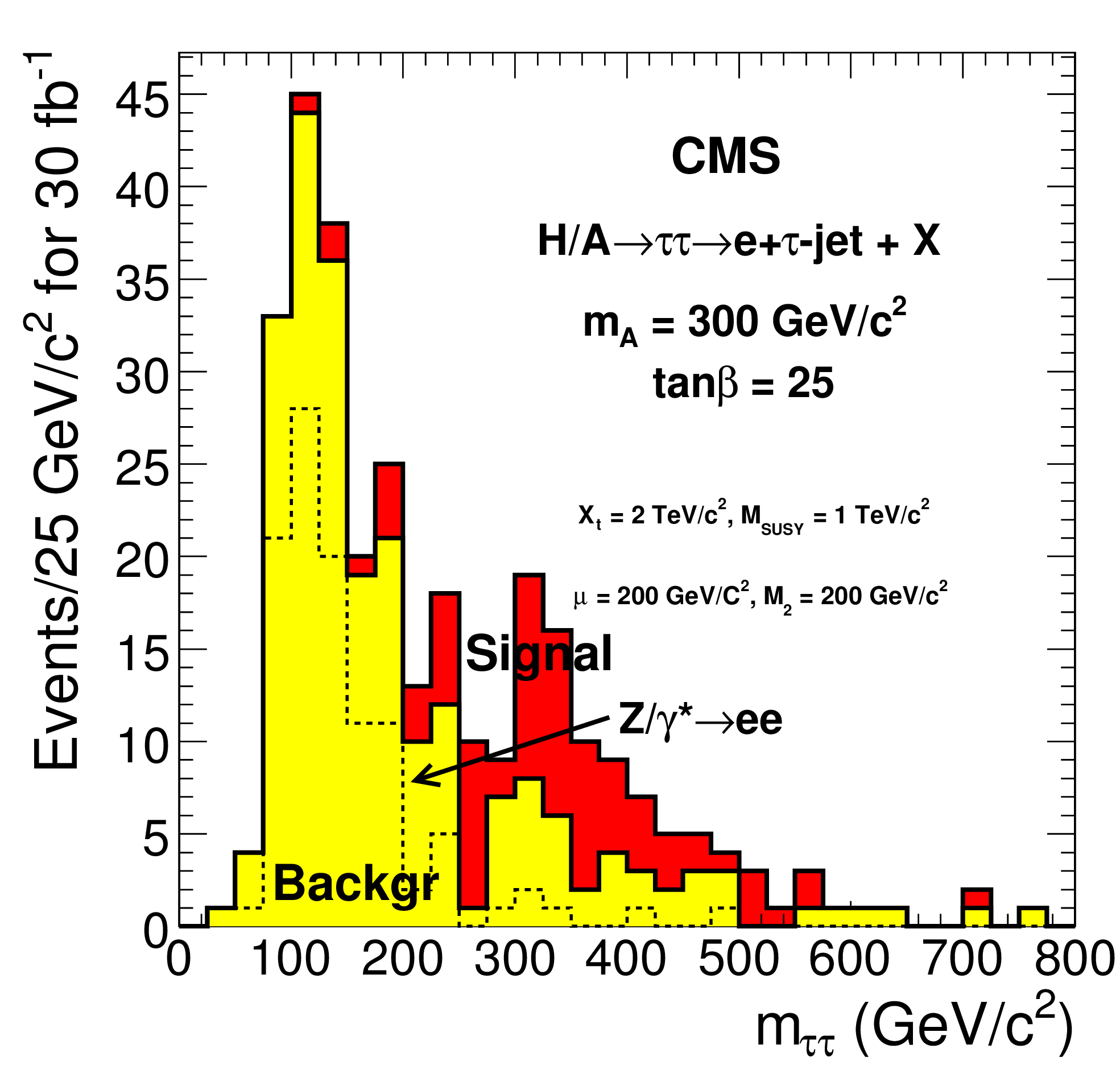}
  \caption{Invariant $m_{\tau\tau}$ distribution for the
    $\mu+\tau\mbox{jet}+X$ (left) and the $e+\tau\mbox{jet}+X$ (right)
    final state for Higgs boson masses, $\tan\beta$ values as
    indicated in the plots, and for an integrated luminosity of
    $30\,\ifb$. In the left plot, the mass windows in which the
    numbers of events are counted for the significance calculation are
    indicated by the vertical arrows.} 
  \label{fig:lhmass}
\end{figure}

\begin{figure}[!t]
  \includegraphics[width=.41\textwidth]{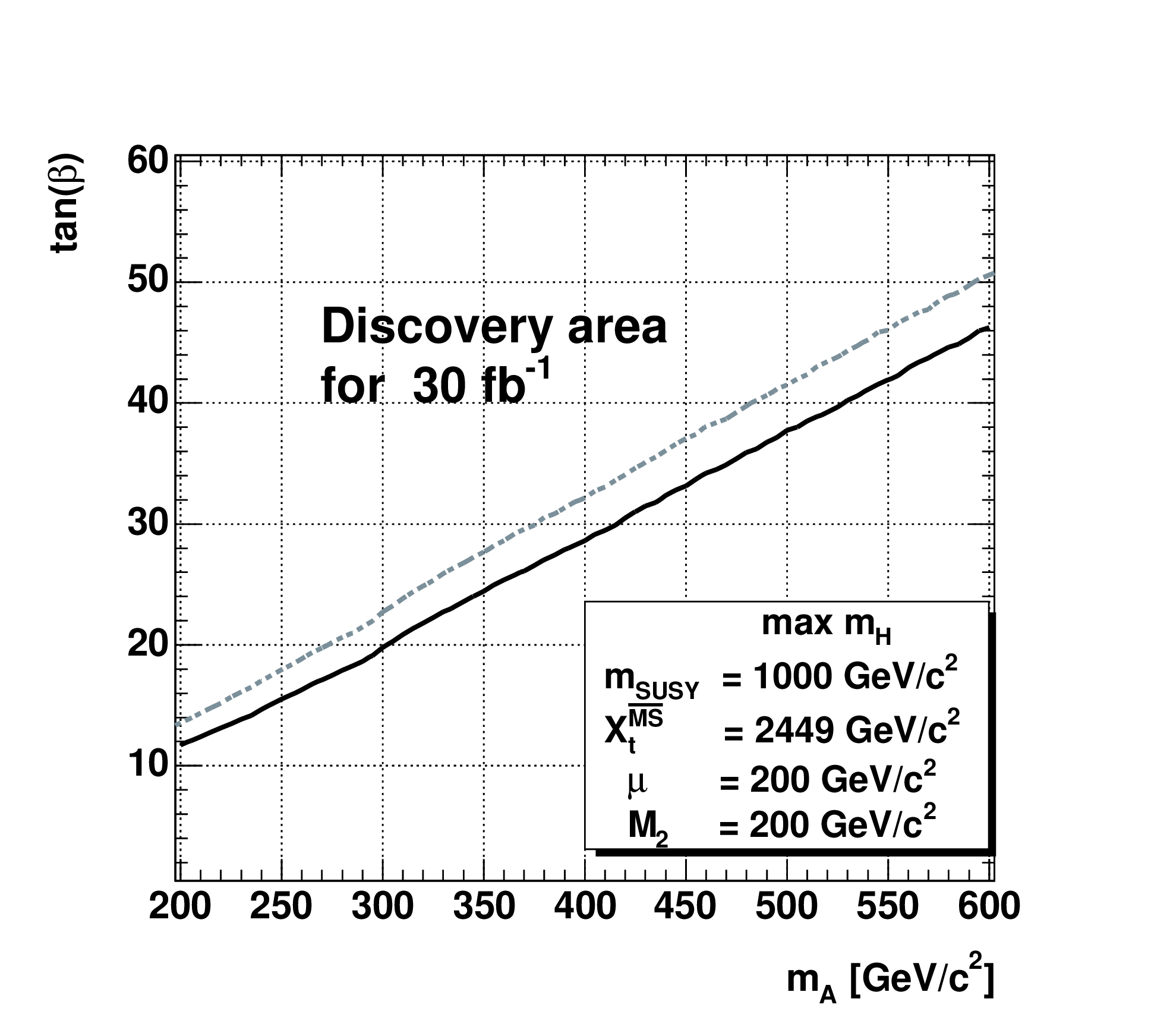}
  \includegraphics[width=.4\textwidth]{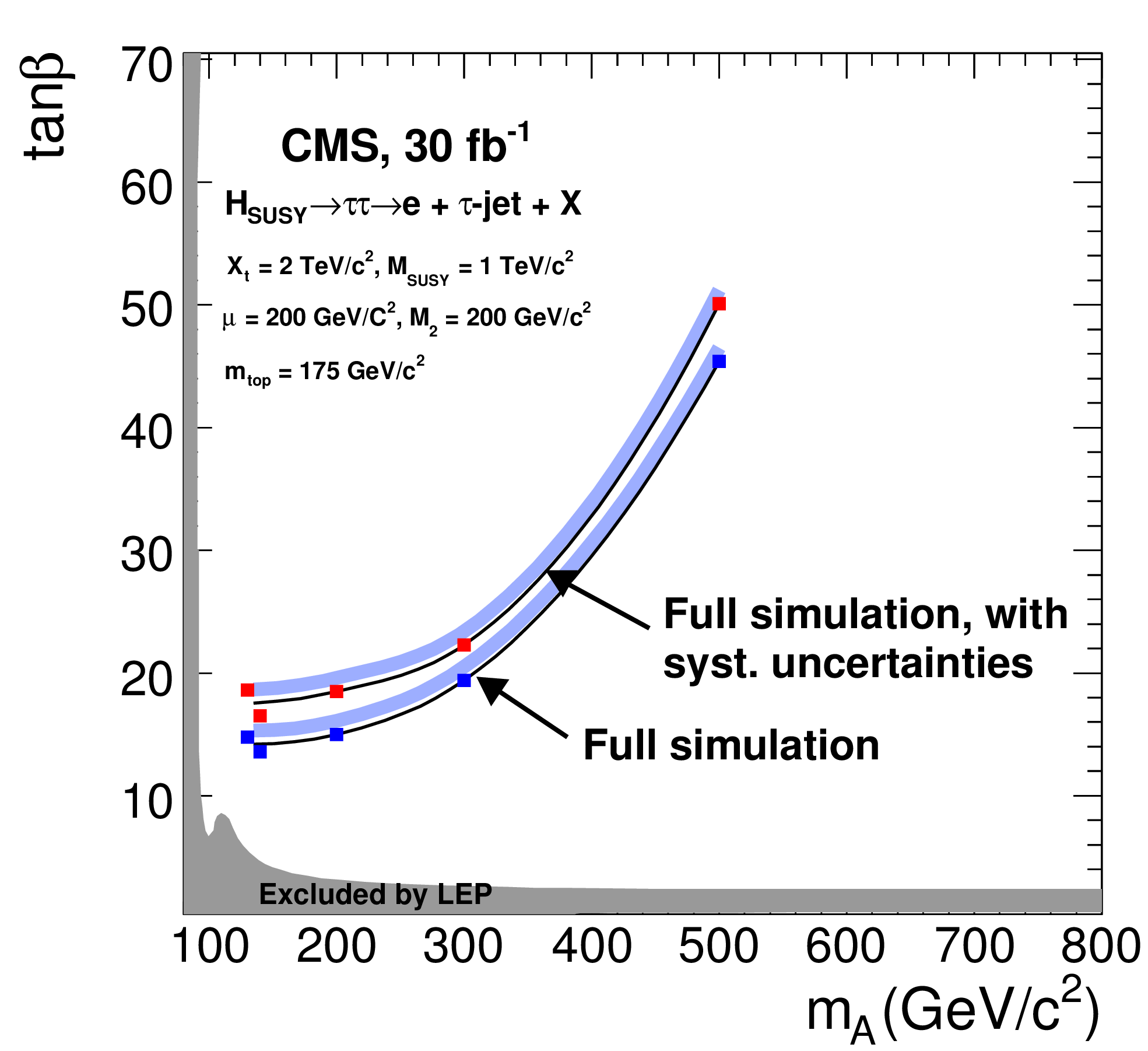}
  \caption{The $5\,\sigma$ discovery contours in the $m_A$
    vs. $\tan\beta$ plane and for an integrated luminosity of
    $30\ifb$. Left: Discovery contour for the $\mu+\tau\mbox{jet}+X$
    final state, with (dashed) and without (solid) the systematic
    uncertainties on the background taken into account. Right: Discovery
    contour for the $e+\tau\mbox{jet}+X$ final state, with and without
    the systematic uncertainties taken into account as indicated in
    the plot. The exclusion limit as obtained from LEP is indicated by
    the shaded area.}    
  \label{fig:lhDiscPot}
\end{figure}

The lepton-hadron final state consists of one electron or muon plus 
jets, one of which is identified as coming from a hadronically decaying $\tau$
lepton, and missing energy in the event. A single electron or muon
trigger, either standalone or combined with a $\tau$ trigger, is used
to preselect the events. One jet in the event is required to be
identified as coming from a $b$ quark in order to suppress backgrounds
from Drell-Yan $\tau\tau$ production, from QCD multi-jet events and
from $W$+jet backgrounds. Details of the analyses can be found in
\cite{CMS:eh,CMS:muh}.  

The invariant $\tau\tau$ mass is reconstructed using the collinear
approximation technique, and is displayed in Figure \ref{fig:lhmass}
for the $\mu+\tau$-jet (left) and $e+\tau$-jet (right) final state,
respectively. 

For the $\mu+\tau$-jet final state the main background after all
selection cuts is 
represented by $\tau\tau b\bar b$, Drell-Yan $Z\to\tau\tau$, and
$t\bar t$ processes. The $t\bar t$ background is estimated from data
by inversion of the electron veto cut and has a systematic uncertainty of
$12.4\%$. The Drell-Yan $\tau\tau$ prediction is taken from a high
precision measurement assumed to be done at the time of this analysis
and a total systematic uncertainty of $8\%$ is assigned. Finally, the
$b\bar b\tau\tau$ background is assumed to be know with a systematic
uncertainty of $15\%$, derived from the uncertainty on a $\mu\mu
b\bar b$ cross section measurement and from the jet-energy scale
uncertainty.  

For the $e+\tau$-jet analysis the main background comes from $(b\bar
b)Z/\gamma^*\to ee/\tau\tau$ final states and from $t\bar t$
processes. The total systematic uncertainty was calculated from the
background uncertainties (either measured or predicted from theory)
and the experimental uncertainties of the event selection, like
electron and $\tau$ identification, calorimetric energy scale and $b$
tagging efficiency. 

The discovery potential for the $\mu+\tau$-jet and the $e+\tau$-jet
analysis is displayed in Figure \ref{fig:lhDiscPot} on the left-hand
side and the right-hand side, respectively. The discovery of a Higgs boson
with a mass of the order of $m_A=200\,\GeV$ would be possible even for
low $\tan\beta$ values, while for high masses such a discovery would
be challenging.

\subsubsection{Higgs Boson Searches in $h/H/A\to\tau\tau\to hh 2\nu$}
This analysis has been performed for Higgs bosons of mass $200$,
$500$, and $800\,\GeV$. The observed final state consists of two
$\tau$-like jets identified by their high transverse energy
and a $p_T>30\,\GeV$ for the leading track inside the
$\tau$-jet. Furthermore, exactly one additional jet with
$E_T>20\,\GeV$ was allowed. This jet had to pass the tagging criteria
for a $b$-jet based on 3D-impact parameters. Details of the analysis
can be found in \cite{CMS:hh}.

After all selection criteria, the dominant background comes from QCD
multi-jet events. In order to estimate this background from Monte
Carlo, the selection has been factorized into three categories:
Trigger and offline calorimetric reconstruction, $\tau$
identification, and finally jet reconstruction, $b$-tagging and
$m_{\tau\tau}$ mass reconstruction. 

The invariant $m_{\tau\tau}$ mass distribution for two Higgs boson
masses ($m_A=200$ and $600\,\GeV$) and for an integrated luminosity of
$60\,\ifb$ are displayed in Figure \ref{fig:hhMass}. The systematic
uncertainties considered are from the $E_T^\mathrm{miss}$ and jet
energy scale ($3-10\%$), tracker misalignment ($\sim 10\%$), and the
measurement of the QCD background from data ($5-20\%$).
Including all systematic uncertainties, a $5\,\sigma$ discovery for a
Higgs boson of mass $m_A=200/500/800\,\GeV$ can be achieved for a
$\tan\beta$ value of $21/34/49$.

\begin{figure}
  \includegraphics[width=.4\textwidth]{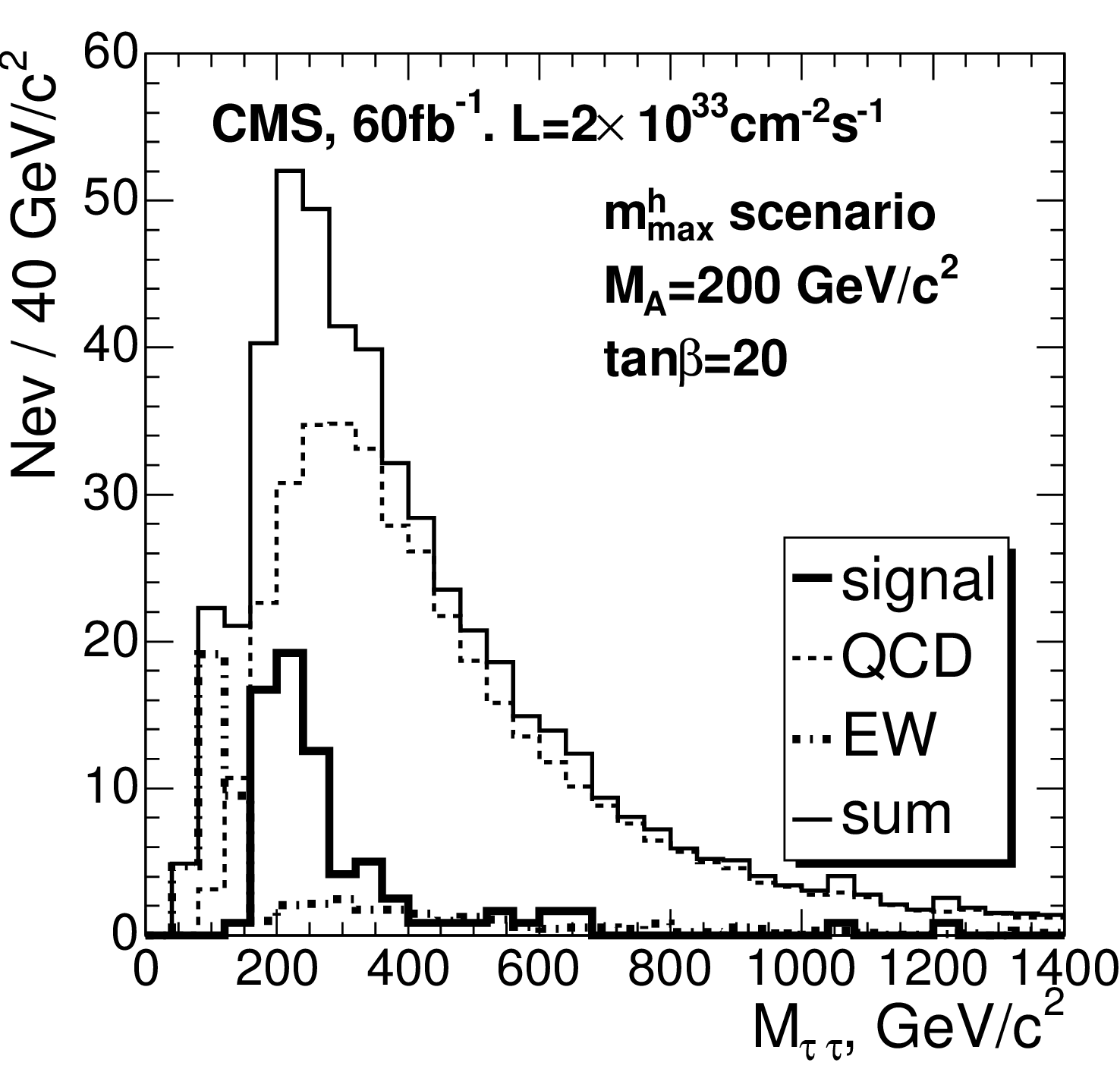}
  \includegraphics[width=.4\textwidth]{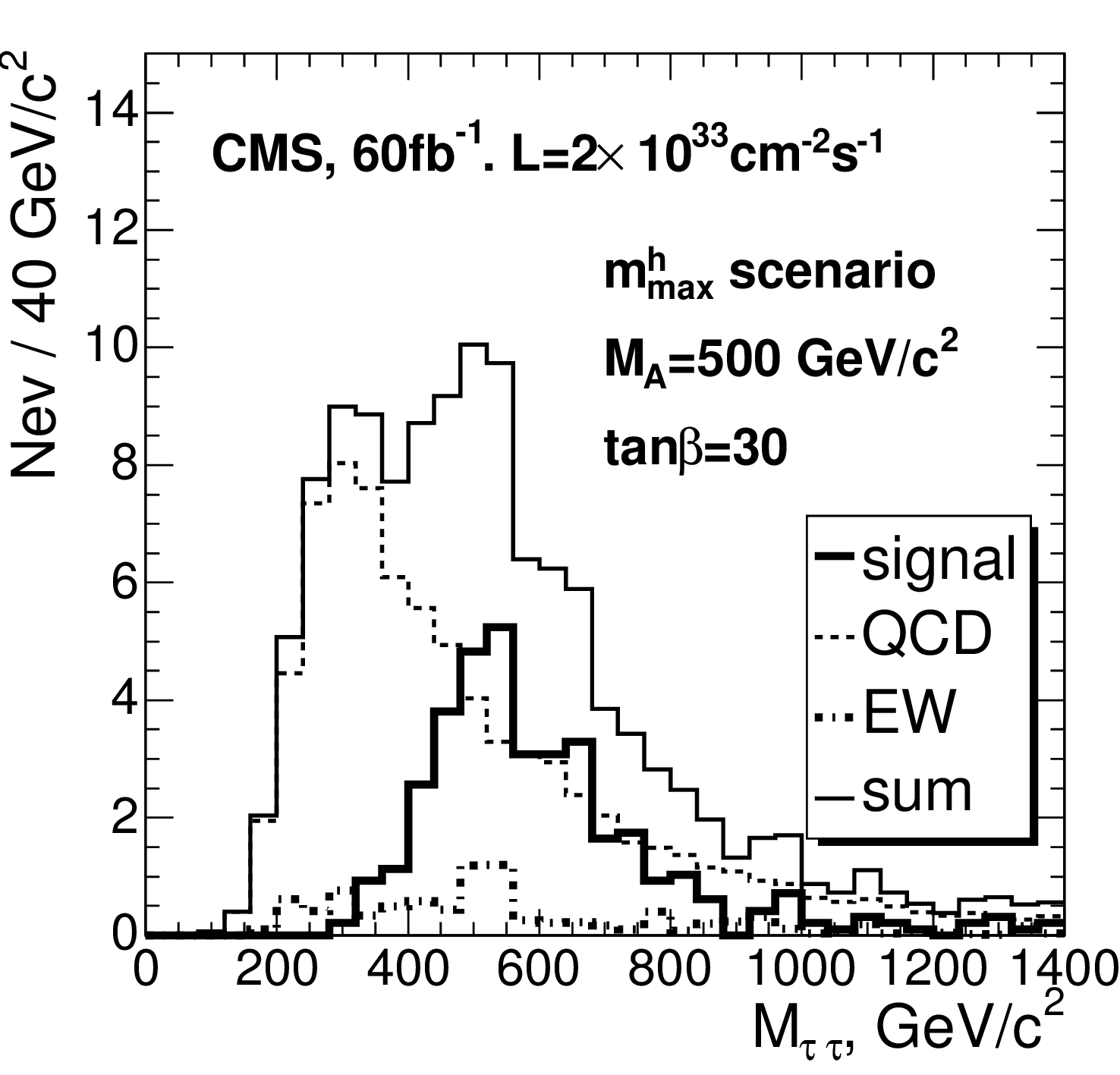}
  \caption{The $m_{\tau\tau}$ distributions for a signal of
    $m_A=200\,\GeV$ with $\tan\beta=20$ (left), for $m_A=500\,\GeV$
    with $\tan\beta=30$ (right), and for the background in $60\,\ifb$
    of data. The solid histogram represents the distribution expected
    from all candidates which is composed of the signal (thick solid
    line), QCD background (dashed line), and irreducible background
    (thick dashed-dotted line).}  
  \label{fig:hhMass}
\end{figure}

\subsection{Searches for the Charged Higgs Boson}
The discovery of a charged Higgs boson would be a definite signal for
the existence of new physics beyond the SM. It is predicted in many
non-minimal Higgs scenarios like Two Higgs Doublet Models (2HDM) or
models with Higgs triplets. The search strategies for a charged Higgs
boson depend on its mass which determines the production rate as well
as the available decay modes. Below the top quark mass, the main
production mode is through top quark decays ($t\to H^+b$) and the
decay of the charged Higgs boson proceeds predominantly via the
$H^+\to\tau\nu$ process. Above the top quark threshold, the production
processes are $gg\to tbH^+$ and $gb\to tH^+$, where the latter
dominates. The decay proceeds predominantly via the $tb$ final state.

Charged Higgs boson searches involve several higher level
reconstructed physics objects such as electrons, muons, jets, jets
tagged as $b$ jets and jets identified as $\tau$ jets. The trigger to
select the relevant event topologies consists of a combination of
$\tau$ triggers, $E_T^\mathrm{miss}$ triggers, and jet
triggers. Details on the analyses can be found in References
\cite{CMS:TDR, CSC:Book}.

\subsubsection{Light Charged Higgs Boson Searches: $m_{H^\pm}<m_t$}
If the charged Higgs boson is light, the branching fraction of
the top quark into a $bW$ final state might be less than that
predicted by the SM. In that case the expected background from
SM-like $t\bar t$ decays is reduced which has to be taken into
account.  

Three different channels have been analyzed, classified according to
the final state of the $\tau$ and $W$ decays:
\begin{itemize}
\item $b\tau(had)\nu\ bW(had)$: This channel has a high branching
  fraction which makes it a priori one of the most important discovery
  channels. However, the absence of leptons and the high hadronic
  activity makes this channel particularly challenging.
\item $b\tau(lep)\nu\ bW(had)$: This channel is characterized by a
  single isolated lepton and large missing $E_T$ due to the neutrinos
  in the final state which make a full kinematic reconstruct of the event
  impossible. The presence of a signal could be detected via the
  excess of kinematically $\tau$-like events or it could be inferred
  from an analysis of the \lq generalized transverse mass\rq\ spectrum
  \cite{Gross:2008ag}. 
\item $b\tau(had)\nu\ bW(lep)$: As above, due to the large number of
  neutrinos in the final state, a full kinematic reconstruction of the
  event will not be possible and a potential signal is again
  recognized by an excess of kinematically $\tau$ like events.
\end{itemize}

The most sensitive channel for a discovery of a light charged Higgs boson is
that where both the $\tau$ and the $W$ decay hadronically. 

\subsubsection{Heavy Charged Higgs Boson Searches: $m_{H^\pm}>m_t$}
For the search for a heavy charged Higgs boson, two channels are
considered:
\begin{itemize}
\item $bqq[b]\tau(had)\nu$: This channel is characterized by one hard
  $\tau$ jet from the decay of the charged Higgs boson, large missing
  transverse momentum, one or two $b$ jets, and two light jets. 
\item $b\ell\nu[b]bqqb$: Here, the charged Higgs boson decays into a
  $tb$ final state. In addition $3-4$ $b$-jets are expected in the
  event (depending on the production mechanism), 2 light quark jets,
  and one high $p_T$ lepton.
\end{itemize}

Of these two channels, only the first one shows a sensitivity to
charged Higgs boson production. 

\subsubsection{Overall Discovery Potential for Charged Higgs Bosons}
In Figure \ref{fig:OverallCH} the overall discovery potential is
displayed for CMS (left) \cite{Ball:2007zza} and
ATLAS\footnote{Here, a statistical uncertainty corresponding to that
  expected for the integrated luminosities given is assumed.}
\cite{CSC:Book} (right). The range of low charged Higgs boson masses up to
the top quark mass is well covered, and a charged Higgs boson can be
discovered with a data set corresponding to an integrated luminosity
of $30\,\ifb$. For charged Higgs boson masses above the top quark
mass, high values of $\tan\beta$ would be necessary for a $5\,\sigma$
discovery.  

\begin{figure}
  \includegraphics[width=.35\textwidth]{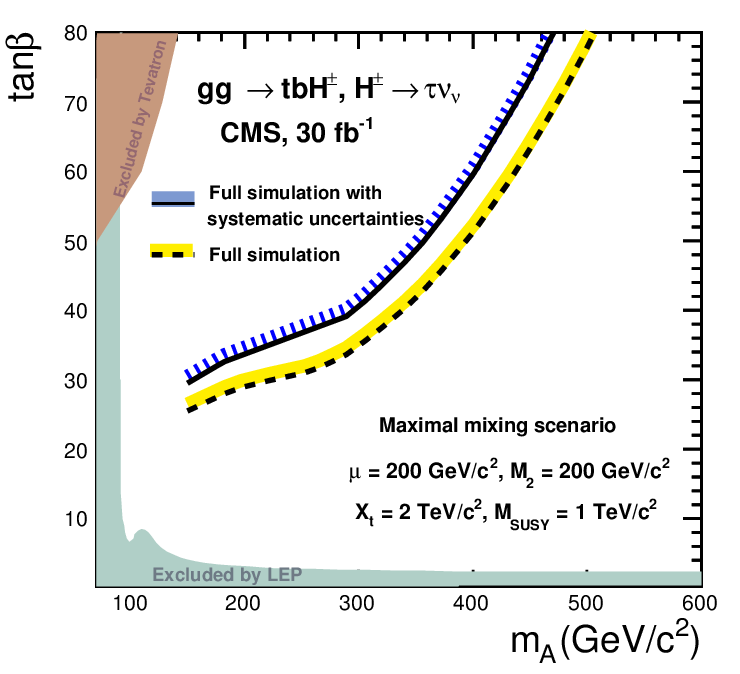}
  \includegraphics[width=.44\textwidth]{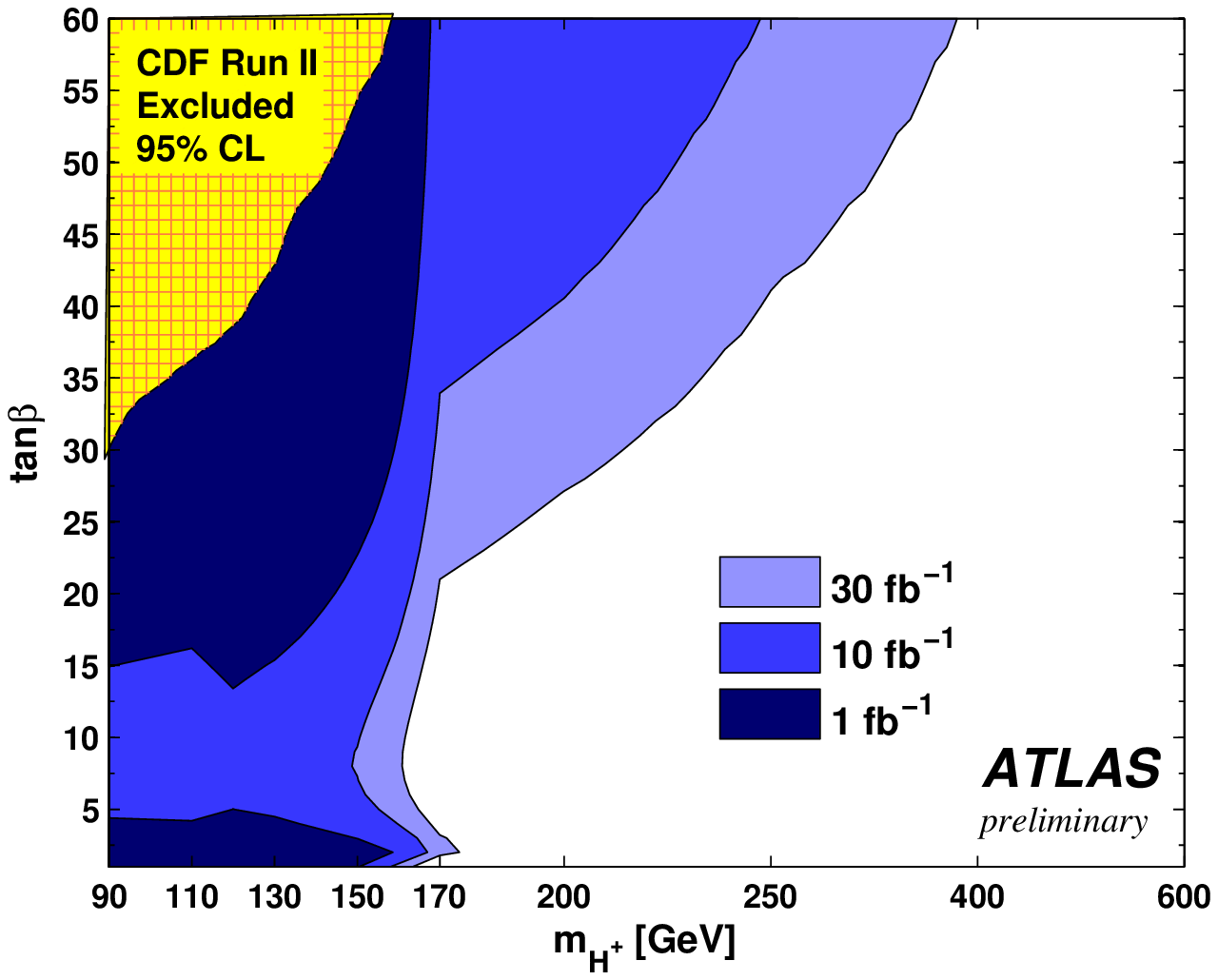}
  \caption{Left: Discovery potential for a charged Higgs boson as a
  function of $m_A$ and $\tan\beta$ for an integrated luminosity of
  $30\,\ifb$ (CMS). Right: Discovery potential for a charged Higgs
  boson as a function of $m_{H^\pm}$ and $\tan\beta$ for integrated
  luminosities between $(1-30)\,\ifb$ (ATLAS). In addition, the $95\%$
  exclusion limit from CDF (Run-II) is shown.}
  \label{fig:OverallCH}
\end{figure}


\section{MEASUREMENT OF HIGGS BOSON PROPERTIES}
Once the Higgs boson has been discovered at the LHC, the measurement
of its properties like mass, spin, CP eigenvalue and its couplings to
fermions and gauge bosons would have to be measured in order to obtain
further insight into the mass generation mechanism realized in nature.

\subsection{The Higgs Boson Width}
In addition to a potential Higgs boson discovery in the
$h/H/A\to\mu\mu$ channel, due to its excellent mass resolution, the
width of the Higgs boson and therefore $\tan\beta$ can be directly
measured \cite{Ball:2007zza}. In Figure \ref{fig:tanbetameasurement}
(left) the intrinsic width of the Higgs boson (circles) and that
measured (solid triangles) for $m_A=150\,\GeV$ is shown. In such an
analysis it has to be taken into account that the mass degeneracy of the
neutral Higgs bosons $A$ and $H$ is not exact which is illustrated by
the open triangles. This is particularly evident for $m_A=150\,\GeV$
and low $\tan\beta$ where the mass difference is larger than the
intrinsic width. In Figure \ref{fig:tanbetameasurement} (right) the
expected uncertainty (including a theoretical uncertainty of $15\%$)
of the $\tan\beta$ measurement is shown as a function of $m_A$ and
$\tan\beta$. The measurement of $\tan\beta$ can be further constrained
by cross section measurements since
$\sigma\times\mathrm{BR}\sim\tan\beta_\mathrm{eff}$.   

\begin{figure}
  \includegraphics[width=.43\textwidth]{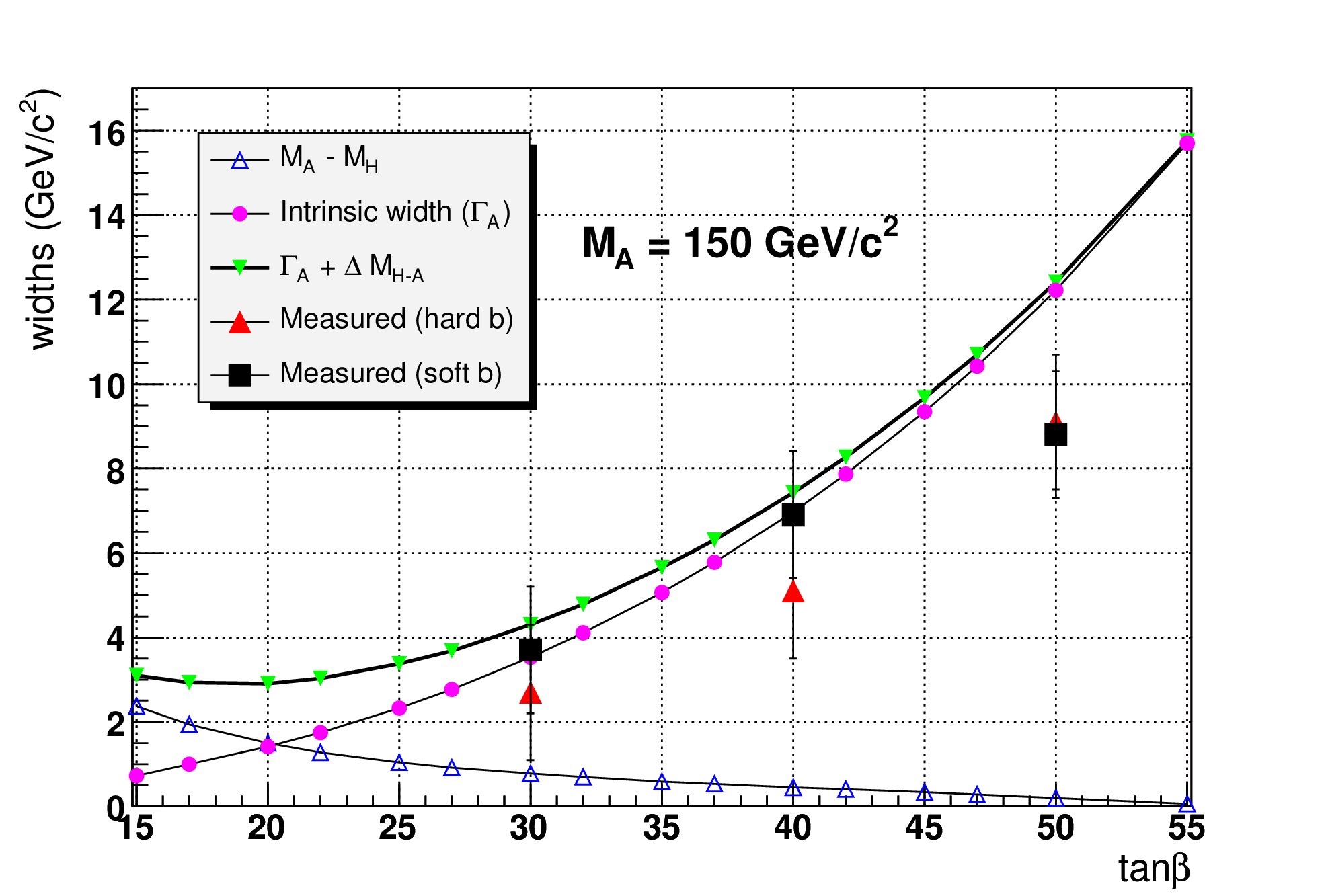}
  \includegraphics[width=.28\textwidth]{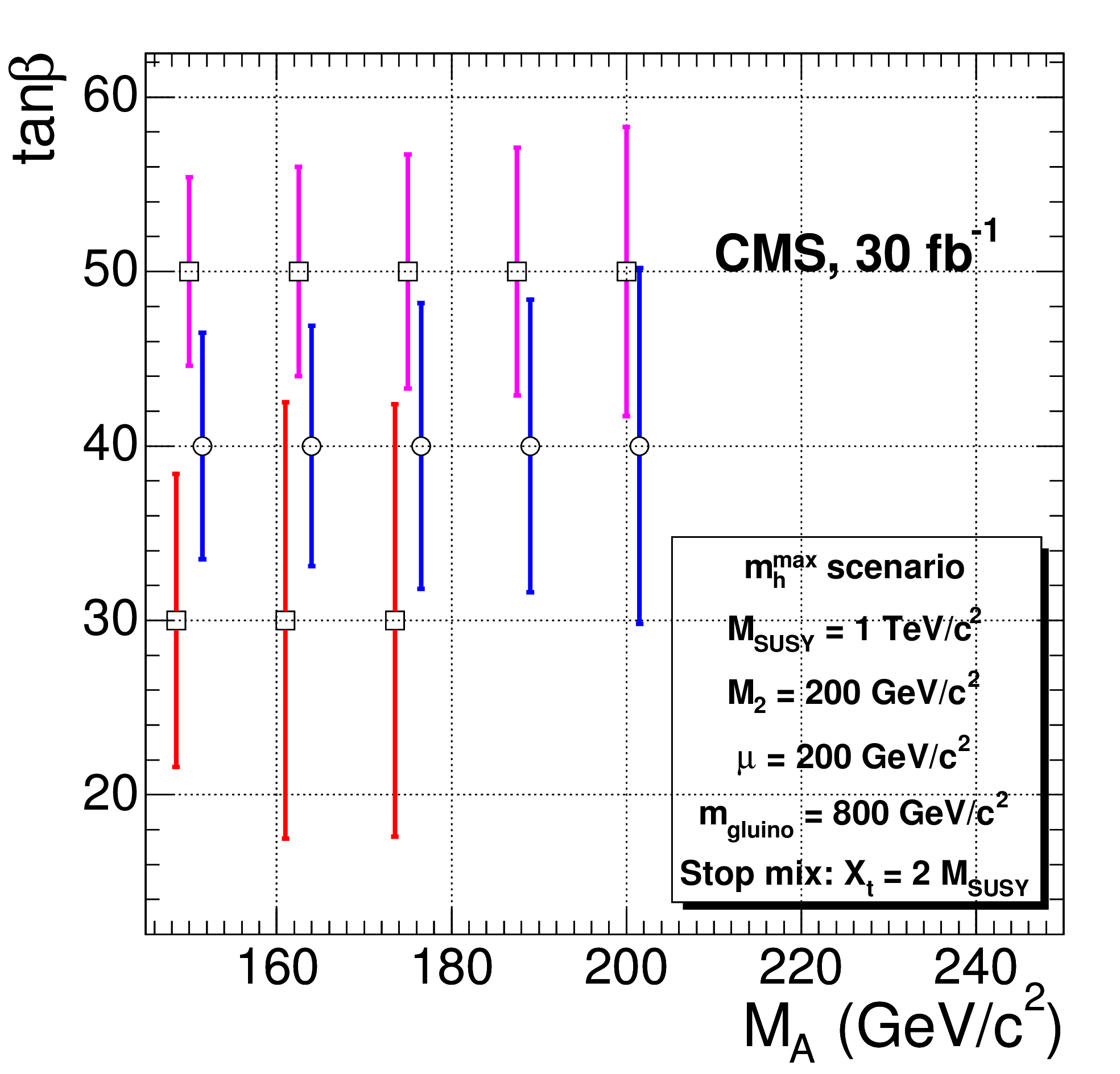}
  \caption{Left: The comparison between the expected Higgs boson width
  and the measured width as a function of $\tan\beta$ for
  $m_A=150\,\GeV$. Right: Uncertainty on the $\tan\beta$ measurement
  obtained from the Higgs boson width measurement with an integrated
  luminosity of $30\,\ifb$.}
  \label{fig:tanbetameasurement}
\end{figure}

\subsection{The Higgs Boson Mass}
In the SM, the mass of the Higgs boson can be measured with the
highest precision in the decay channels $H\to ZZ^{(*)}\to 4\ell$, in
$H\to\gamma\gamma$, and in $H\to b\bar b$ \cite{ATLAS:TDR}. However,
recent studies indicate that a discovery of the Higgs boson in the
$b\bar b$ and therefore a measurement of its mass might be
challenging \cite{CSC:Book}. Assuming $300\,\ifb$ of data, a relative
precision of such a measurement of $0.1\%$ in the range
$100<m_H<400\,\GeV$ can be achieved as displayed in Figure
\ref{fig:HiggsMass} (left). The anticipated precision degrades down to
$1\%$ for Higgs boson masses of $m_H=700\,\GeV$. The systematic
uncertainties are dominated by the knowledge of the absolute
calorimetric energy scale which, for leptons and photons, was assumed
to be $0.1\%$. However, a final value of $0.02\%$ for that uncertainty
is anticipated driven by the objective to measure the mass of the $W$
boson with a precision of $15\,\MeV$. An uncertainty of $1\%$ on the
jet energy scale was assumed. 

\begin{figure}[!b]
  \includegraphics[width=.41\textwidth]{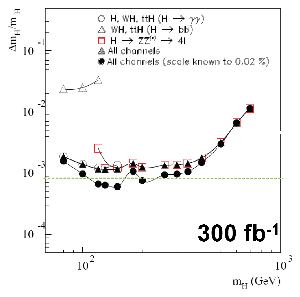}
  \includegraphics[width=.38\textwidth]{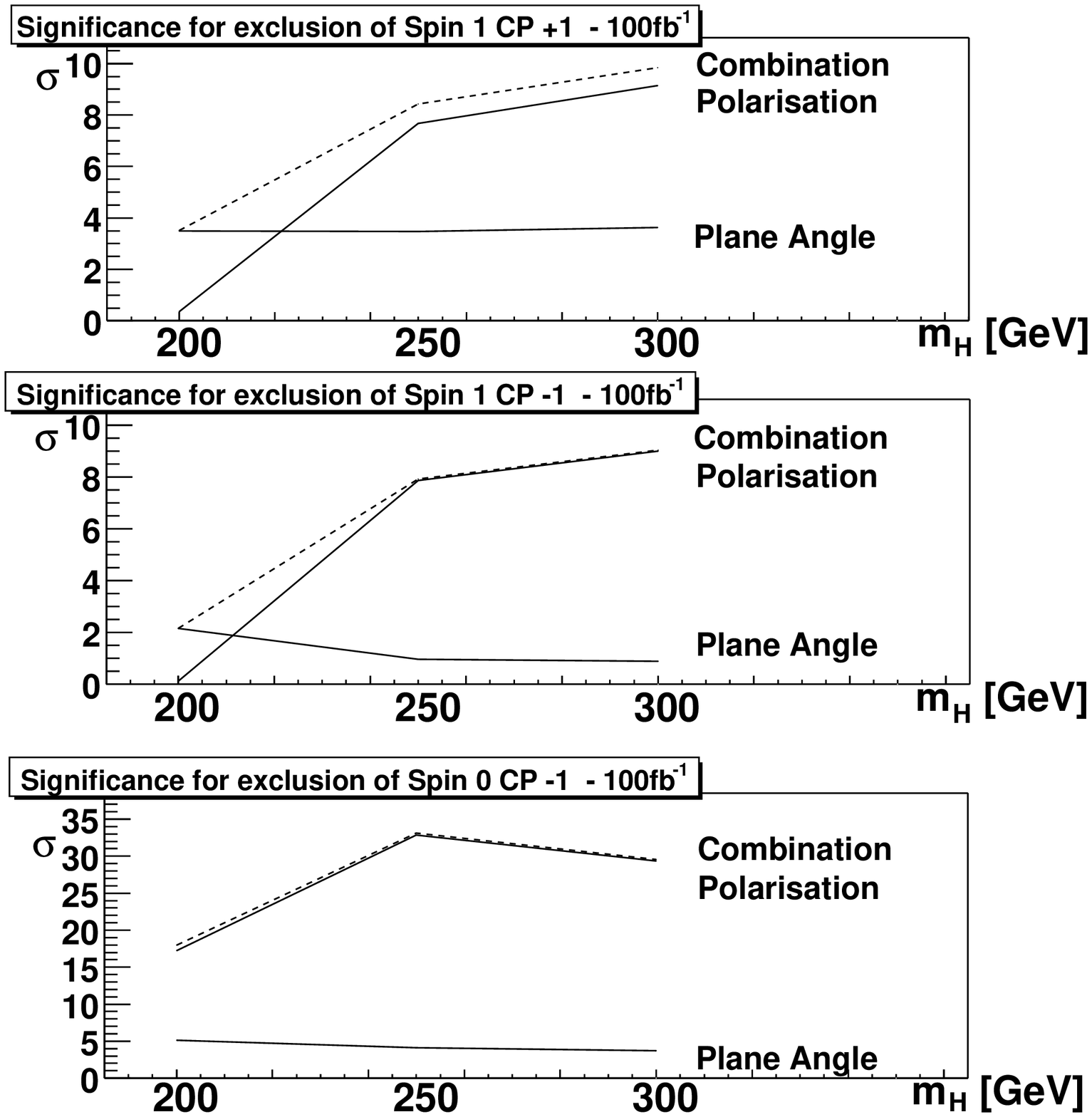}
  \caption{Left: The anticipated precision of a measurement of the Higgs
    boson mass in the decay channels $H\to\gamma\gamma$ (open
    circles), $H\to b\bar b$ (open triangles), and $H\to ZZ^{(*)}\to
    4\ell$ (rectangles). A precision of the order of $0.1\%$ can be
    achieved for Higgs boson masses of $100<m_H<400\,\GeV$, while for
    $m_H=700\,\GeV$ $1\%$ is expected in $300\,\ifb$ of data. Right:
    The significance for the exclusion of a Higgs boson with non-SM
    like spin and CP eigenvalues as a function of its mass.} 
  \label{fig:HiggsMass}
\end{figure}

In models beyond the SM (like the MSSM), the coupling of the Higgs boson
to gauge bosons might be suppressed or even absent. In that case
the mass of the Higgs boson can be measured, e.g. in the $H\to b\bar
b$, $H\to\mu\mu$ or $H\to\tau\tau$ channel. In the latter case, due to
the neutrinos from the $\tau$ decays, one would have to use techniques
like the collinear approximation to reconstruct the mass of the Higgs
boson. In all cases, a precision on the $\%$ level or better can be
expected.

\subsection{Spin and CP of the Higgs Boson}
After the discovery of the Higgs boson, the highest priority is to
establish its spin and CP eigenvalue. This can be determined
from a study of the angular distributions and correlations in the
$H\to ZZ{(*)}\to 4\ell$ channel \cite{Buszello:2002uu,Bluj:2006}. The
angular distributions investigated are the polar angle of the leptons
relative to the direction of flight of the $Z$ boson, and the angle
between the decay planes of the two $Z$ bosons, both calculated in the
Higgs boson rest frame.

The result of this study is shown in Figure \ref{fig:HiggsMass}
(right) for non-SM like spin and CP configurations for the Higgs
boson. For masses larger than $230\,\GeV$ a spin-1 hypotheses with
either even or odd CP eigenvalue can be ruled out with a data set of
$100\,\ifb$. For masses as low as $200\,\GeV$ and a
luminosity of $300\,\ifb$, a Higgs boson with (spin, CP)=(1,+) can be
ruled out at the $6.4\,\sigma$ level, while for (spin, CP)=(1,--) a
significance of only $3.9\,\sigma$ can be obtained. A Higgs boson with
(spin, CP)=(0,--) can be ruled out with less than $100\,\ifb$ over the
whole mass range considered here.

\subsection{The Higgs Boson Couplings}

\begin{figure}[!b]
  \includegraphics[width=.27\textwidth]{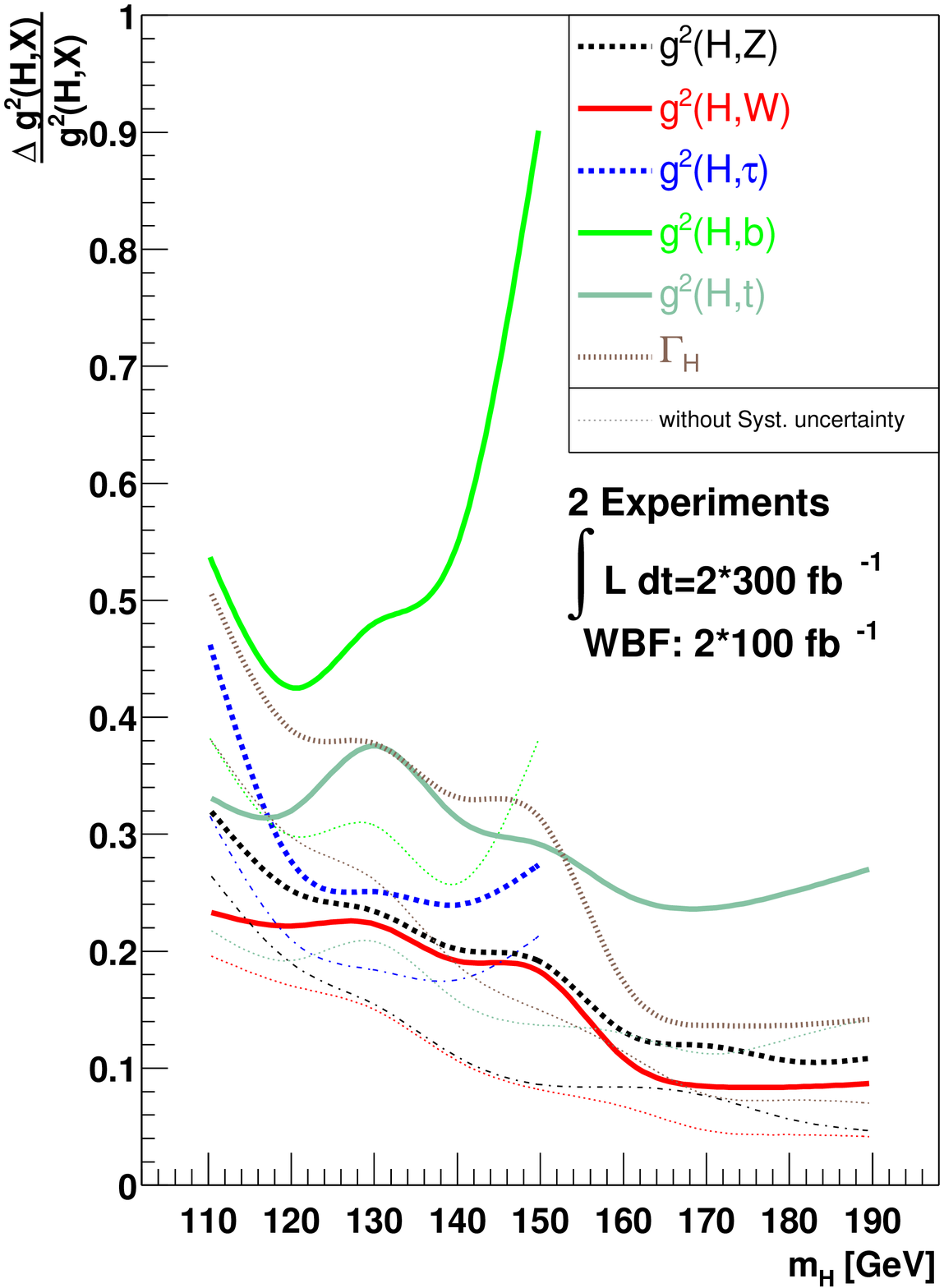}
  \includegraphics[width=.27\textwidth]{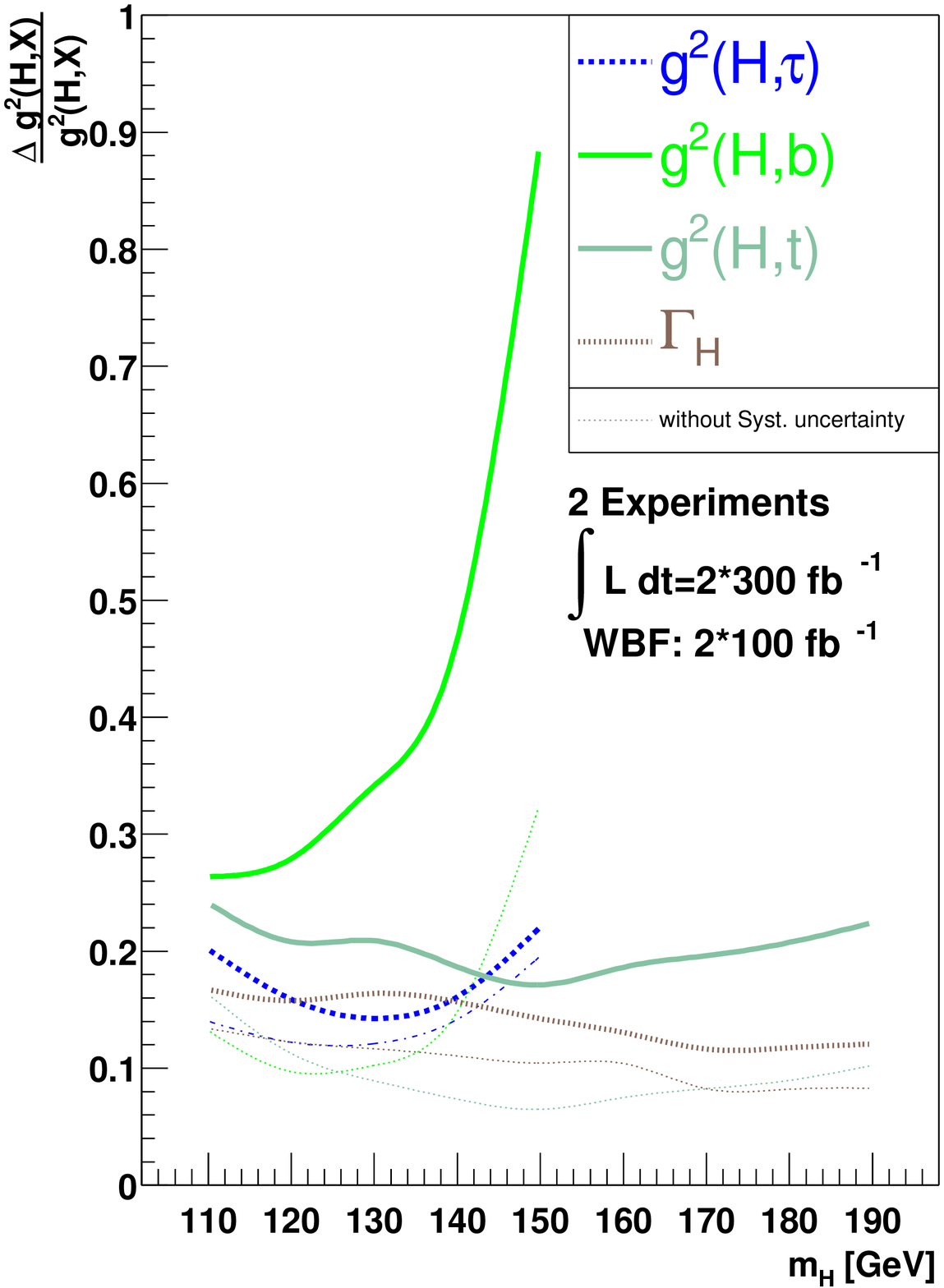}
  \caption{Relative precision of a measurement of the Higgs boson
  couplings squared as a function of $m_H$ for an integrated
  luminosity of $300\,\ifb$ per experiment with moderate (left) and
  more restrictive (right) theoretical assumptions as explained in the
  text.}
  \label{fig:HiggsCouplings}
\end{figure}

The coupling of the Higgs boson to gauge bosons and fermions
determines its production cross section and branching
fractions. Measuring the rates in multiple channels allows for a
determination of various combinations of couplings at the LHC
\cite{Duhrssen:2004cv}. However, a model independent measurement of
the (partial) decay width(s) of the Higgs boson will not be possible
at the LHC. The reason being that on the one hand a measurement of the
missing mass spectrum like at $e^+e^-$ colliders is not possible, and
that on the other hand some of the decay channels of the Higgs boson
are either hard (like $H\to b\bar b$) or even impossible (like $H\to
gg$) to detect due to the overwhelming background from QCD
processes. An absolute measurement of the above quantities will only
be possible if additional theoretical assumptions are being made.
Introducing moderate theoretical constraints as detailed below, allows
to overcome the experimental difficulties described above and to
measure the couplings of the Higgs boson.

In a first approach, only the couplings of the Higgs boson to the gauge
bosons is constrained to less than 1.05 times its value in the
SM\footnote{The additional $5\%$ account for theoretical uncertainties
  in the translation between the couplings squared and the partial
  widths, and also for small possible admixtures of exotic Higgs states.}, which
is justified in any model with an arbitrary number of Higgs
doublets. Furthermore, additional particles running in loops in the
$H\to\gamma\gamma$ and $gg\to H$ processes are allowed. The relative
precision of the measured couplings as a function of the Higgs boson
mass is illustrated in Figure \ref{fig:HiggsCouplings} (left) assuming an
integrated luminosity of $300\,\ifb$ per experiment. The couplings of
the Higgs boson to the $W$ and $Z$ bosons, to the top quark, and to
the $\tau$ lepton can be measured with a precision between $(10-40)\%$
depending on $m_H$. 

If in addition, the ratio of the $W$ and $Z$ couplings to the Higgs
boson are constrained to within $1\%$ of that in the SM, the absolute
value of the $W$ coupling to within $5\%$, and no new non-SM particles
are allowed in the loops, a higher precision can be obtained as
illustrated in Figure \ref{fig:HiggsCouplings} (right). In that case a
precision of the measurement of the couplings squared can be achieved
which is about a factor of two higher.
 
\section{SUMMARY}
In this note, the discovery potential for neutral ($h/H/A$) and charged
($H^\pm$) Higgs bosons in the MSSM in ATLAS and CMS at the LHC has
been reviewed. After a potential discovery of a Higgs boson, its
properties have to be measured in order to gain insight into the mass
generation mechanism realized in nature. Possible measurements of the
Higgs boson mass and width, its spin and CP eigenvalue, and its couplings to
fermions and gauge bosons have been discussed. 

\begin{acknowledgments}
The author wishes to thank the organizers of the conference for the
invitation and everybody in the ATLAS and CMS Higgs working groups
that helped in preparing the talk.
\end{acknowledgments}

\end{document}